# Beam modelling of Hitachi PROBEAT proton therapy system for a GPU-based Fast Monte Carlo dose engine

**Qianxia Wang[1, 2, 3*], Poenisch Falk[3], Yao Zhao[3], Xueming Bai[4], Roelf Slopsema[5], Kirk Jon Luca[5], Thomas J Whitaker[3], Yun Hu[6], Uwe Titt[3], Radhe Mohan[3], Pablo Yepes[2, 3*]**

[1]Department of Radiation Oncology, University of Nebraska Omaha, 6001 Dodge St, Omaha, NE 68182, USA
[2]Department of Physics and Astronomy, MS 315, Rice University, 6100 Main Street, Houston, TX 77005, USA
[3]Department of Radiation Physics, Unit 1420, The University of Texas MD Anderson Cancer, 1515 Holcombe Blvd., Houston, TX 77030, USA
[4]Mevion Medical System, Kunshan, Jiangsu, China
[5]Department of Radiation Oncology and Winship Cancer Institute, Emory University, Atlanta, GA 30308, USA
[6]Division of Radiation Oncology, The University of Texas MD Anderson Cancer, 6565 MD Anderson Boulevard, Houston, Texas, 77030, USA

*E-mail: yepes@rice.edu and qzw0004@auburn.edu

### Abstract

Background: An in-house dose engine independent of clinic TPS is not only a reliable tool for patient QA verification. More importantly, it plays vital role in cutting-edge research due to its flexibility in implementing new functions. In this study, we upgraded our existing beam model with using double-Gaussian distributions for both spatial and opening angle distributions of particles to obtain more accurate phase space files. It is expected to potentially improve the performance of this independent dose engine in both clinic and research at the expanded MD Anderson proton center.

At the same time, we changed the openning angle distribution center from along z-axis to the particle's original direction which complicated the simulation but is more reasonable. All lateral dose profiles were significantly improved after the beam model updates from the gamma-index passing rates (from [70.31%, 100%] presented in AAPM 2020 abstract to [100%, 100%]).

The new generated phase spaces files were further verified with 2744 IMPT QA measurements. At the same time, results from a widely used TPS (RayStation) were also presented and compared with our in-house dose engine.

Objective: To model Hitachi PROBEAT IMPT beam for a GPU-based Fast Monte Carlo dose engine, the Fast Dose Calculator (FDC) and verify its accuracy with different IMPT QA plans.

Approach: Double-Gaussian distributions were applied to accurately model spatial and angular distributions of particles in Phase Space (PS) plane located 490 mm up-stream from the isocenter. A total of 12 parameters were used for each phase space generation. To validate the model, simulated integral depth dose (IDD) in water and lateral profiles at five depths (- 200, -100, 0, 100 and 200 mm) in air were compared with measurements for 20 single-energy beamlets. Particle numbers per monitor unit (MU) for these 20 energies were derived from absolute dosimetric measurements. PS parameters and particle numbers per MU for the remaining of available energies were linearly interpolated from the modelled energies. The generated PS and MU files were subsequently evaluated on 269 IMPT plans. FDC and TPS doses were compared with measurements. Quantitative evaluation was performed using 1D (1%/1 mm for commissioning data) and 2D (3%/3 mm for IMPT plan QAs) gamma-index analysis.

Main results: Gamma-index passing rates were ≥ 97.6% for IDD and 100% for all lateral profiles. For the 2744 measurements at various depths from IMPT QAs, mean, median [range] of gamma-index passing rates were 98.5%, 99.8%, [41.8%-100%] for FDC and 98.7%, 100% [39.4%-100%] for TPS. TPS is 0.2% better than FDC in both mean and median. Across different treatment sites, mean is with a range of 95.0%-99.5% for TPS and 95.8%-99.3% for FDC. FDC is 0.2% and 0.8% higher in mean for H&N and breast.
Conclusion: High gamma-index passing rates for IDD and lateral profiles demonstrate accurate commissioning of FDC. Large-scale comparison between IMPT QA measurements and calculation further confirms that the commissioned FDC is sufficient accurate for clinic implementation.

---

## 1. Introduction

An increasing number of patients prefered Proton therapy due to its superiority in the deep penetration and reduced exit dose compared with conventional photon and electron therapies[1]. The University of Texas MD Anderson Cancer Center expanded its Proton Therapy Center to meet the high demand for proton treatment service and enable patients' access to the most advanced technology[2]. This expansion includes more than doubled the center size, a total of 8 proton treatment rooms and two synchrotrons[3].

An in-house dose engine independent of clinic TPS is not only a reliable tool for patient dose reverification. More importantly, it plays vital role in cutting-edge research due to its flexibility in implementing new functions, such as updating existed models, adding new quatity calulations, new beam shaping apetures, etc. The Fast Monte Carlo Calculator (FDC)[4,5], a track-repeating Monte Carlo Algorithm, has been sucessfully utilized as an independent dose engine for patient specific QA simulation. As a research tool, it has played key roles in different projects, such as LET optimization[6], a novel RBE model[7], variable and constant RBE comparision[8], proton arc[9], carbon therapy[10], flash proton beam shaping and optimization[11], etc.

To enable the FDC for a new proton machine, the first step is to accurately model mono-energetic beamlets according to measured commissioning data and estimate the number of particles per monitor unit (MU) according to absolute dosimetry measurements. The phase space files[12] used by the on-going FDC for the MD Anderson Proton Center were from beam modelling by the MCNPX[13], another Monte Carlo dose engine used for research at MD Anderson. Later when using FDC for Shanghai Proton and Heavy Ion Center (SPHIC), we developed our own beam model[14,15], for which dose energy, spatial and opening angle distributions were described by single Gaussian functions. The single Gaussian distribution was centered at real energy which was close to the nominal beamlet energy. The spatial and opening angle distributions were centered at 0 which means most of particles were located at the center of the x-y plane and travelled along z axis.

With this beam model, FDC simulation is unable to match the measured lateral profile halo[16,17] for the new proton synchrotron at the Proton Therapy Center 2 MD Anderson proton center. To improve the simulation, we updated the existing beam model using the double Gaussian functions to fit spatial and opening angle distributions. Another modification to the existing model is the open angle distribution center, which was from along z axis to a proton's original direction (along the line from virtual source to the location in the phase space x-y plane).

To further assess the accuracy of the commisioned FDC, we applied the newly generated phase space and MU files to 269 IMPT QA plans. A gamma-index analysis[18] was performed to compare FDC calculated dosess with measurements at different depths for each beam within the QA plans. Raystation is a TPS that is widely used in clinic for proton threapy[19]. For comparison, we calculated the same IMPT QA plans with RayStation and evaluate the agreement between calculated dose and measured dose by gamma-index.

## 2. Materials and Methods

### *2.1 Commission data measurement and equipment*

Comminissioning data, including the IDD in water and lateral profile in air, is required to model the beam for a dose engine. The IDD data were aqcuired by combining water tank scan using a Bragg-peak chamber with the IBA Giraffe muti-layer ion chamber. The Giraffe data (2 mm electrode separation, 12 cm diameter and 180 electrodes) were used in the proximal region (less than 26 mm depth) and the Bragg peak chamber (diameter of 8 cm) was used at deeper depth. Both data fit to regularize the spacing to 0.25 mm using Python PyBragg function. The lateral profile was measured in air with an amorphus silicon detector IBA Phoenix with a resolution of 0.2 mm in both axis directions. Twenty energies were measured with a range of 70.2 MeV - 228.7 MeV (Table 1).

The IDD and lateral profile data provides the shape of the dose distributions. In addition, an absolute dose measurement is required for each beam to determine the absolute dose or number of protons per monitor unit. PTW Roost Type 34001 with a lateral diameter of 15.6 mm was used for the absolute dose measurement at different depths for different energies. As the energy increases, the measurement depth increases from 14 mm to 87 mm. The absolute dose for a single energy was measured with a delivery plan that has a field size of 100 mm × 100 mm, 2.5 mm spot spacing and uniform spot weights. Including the 20 energies mentioned above for the commissioning data, another 4 extra energies (73.9 MeV, 84.3 MeV, 93.9 MeV and 102.7 MeV) were also included for the absolute dose measurement.

### *2.2 Beam modelling*

The phase space plane perpendicular to the beam direction (z axis) was selected at 490 mm upstream of the isocenter, and it is described with 12 variables. The energy distribution was assumed to be single gaussian with 2 variables, mean energy ($E_r$) and FWHM ($w_E$). The spatial distribution of particles in this plane was defined by two double gaussian distributions centered around 0 for x and y axes respectively with 6 variables, the FWHM ($w_{1x/y}$) for the 1st gaussian, the FWHM ($w_{2x/y}$) for the 2nd one and the relative weight of the second gaussian to the first one ($We_{x/y}$).

Cosine of the projected angles to x and y axes (cos x and cos y) were used to describe proton's opening angle relative to the z-axis at the phase-space plane. The distributions of cos x and cos y were also modeled with double-Gaussian functions. For each proton, the center of this angular distribution was defined by its original direction, i.e., the line from the source to the proton's physical location on the phase-space plane. Because different protons intersect the plane at different positions, their original directions vary, which causes the opening-angle distribution to differ from proton to proton.

To address this, we introduced a temporary Cartesian coordinate system (x′, y′, z′) for each proton, where the z′ axis is aligned with that proton's original direction. In this temporary coordinate system, the opening angle can be sampled using two uniform double-Gaussian distributions—one for the x′-projection and one for the y′-projection. After sampling the opening angle in the temporary coordinate system, the resulting direction was transformed back into the original coordinate system and stored in the phase-space file.

Four additional parameters were used to describe the angular distribution: the FWHM values $w_{1cosx}$ and $w_{1cosy}$ for the first Gaussian, the FWHM values $w_{2cosx}$ and $w_{2cosy}$ for the second Gaussian, and the relative weight of the second Gaussian, which was taken to be the same as the spatial-distribution weight Wex/y.

### *2.3 Proton number per MU*

To determine the number of protons per MU for each single energy, we first calculated the dose distribution per proton with the generated phase space file. The calculation is based on the delivery plan and a water phantom with a dimension of 150 mm × 150 mm × 100 mm, with the isocenter set at the tank entry.

Then we scaled the calculated point dose to the measured absolute dose per MU. This scale factor was the number of protons per MU for the specific energy. With this method, 24 scale factors were determined for the 24 energies that have absolute dose measurements. Proton numbers per MU for other energies available for the machine were interpolated from values of these 24 energies.

### *2.4 IMPT plan and QA measurement*

A total of 269 IMPT plans for 155 patients were selected for the commissioned dose algorithm verification. These IMPT plans included different treatment sites. For each plan, there are 1-5 beams in them. Measurements were performed for each beam at different depths ranging from 6 cm to 290 cm. All 269 IMPT plans were calculated independently by both FDC and RayStation, enabling a paired comparison of the two algorithms against the same set of measurements.

We divided all these plans into 10 categories: brain & skull base, CNS, spine, H&N, breast, lung & other thoracic cancer, GI, pelvic, prostate and other (skin, skull, thigh and gluteus). The number of patients, plans, beams and measurements for each category are show in Table 2.

The equipment used for the IMPT QA measurement is IBA MatriXX PT, an ionization chamber detector array with a pixel spacing of 7.6 mm. There are 1020 ion chambers with a measurement area of 24.4 × 24.4 $cm^2$.

## 3. Result

### *3.1 Commissioning data*

The 12 variables for each phase space file were determined by matching FDC simulation with commissioning data. A totoal of 20 energies were provided commisioning data and their phase space files were generated one by one by tuninng variables to match measurements. Figure 1 displays the comparison of IDD calculated by FDC (black solid lines) and measured (red dotted lines) for 20 energies. The gamma-index passing rate ranges from 97.6% to 100% with the criterion of 1%/1 mm (Table 1).

Comparing with higher energies, the passing rates for low energies were slightly lower and the deviations lied at shallow depth. The first three energies have a passing rate below 99%. The deviation at the entrance is around 4.4% for 70.2 MeV, 5.1% for 79.2 MeV and 5.3% for 89.2 MeV. The deviation was for the integrated dose over the x-y plane, a smaller difference was expected for point doses in the plane.

Figure 2 shows the comparison of lateral profiles simulated by FDC (black solid lines) and measured (red dotted lines) for 4 of 20 energies at 5 depths (-200 mm, -100 mm, 0 mm, 100 mm and 200 mm). Plots of the left column are presented in linear scale, and those on the right present the results in log scale in order to display comparison details at very low dose ($10^{-4}$). Plots and high gamma-index passing rates (100%) in Table 1 indicate the good agreement.

### *3.2 Absolute dose calibration*

Per TG185[20], the absolute dose calibration is recommended to be performed at least 1 cm depth at a relatively flat plateau region. Unlike some other proton centers using the same depth for measuring absolute dose of various energies, different depths (14 mm - 87 mm) were selected for a flatter plateau region for different energies in this new proton center.

Figure 3 shows the number of protons per MU for different energies. The number of protons per MU increases as the beam energy increases. And the increase rate decreases for the higher energy region. Because protons with high energy travel further distance and deposit less dose in the shallow region compared with low energy protons. It requires more protons to accumulate the same amount of dose at the flatter plateau region.

The red triangles in Figure 3 corresponded to values calculated from the measured absolute doses for the 24 energies. Black spots represent interpolated values.

### 3.3 IMPT plan QA verification

With the phase space and number of protons per MU for different energies, we calculated patient IMPT QA and compared with measurements for verification. A total of 2744 measurements were collected for FDC verification (Figure 4, 5 and Table 2). RayStation is a clinically established TPS for proton therapy. To enable a direct paired comparison, the same 2744 measurement planes used for FDC verification were also compared with RayStation calculations of the identical plans.

The gamma-index passing rate of different measurements comparing with FDC is between 41.8% and 100% with the criterion of 3 mm/3%. The mean value of passing rates is 98.5% and the median value is 99.8%. The range for gamma-index passing rates for measurements comparing with TPS is 39.4% - 100% with the same criterion. The mean value is 98.7% and the median is 100%.

Analysis was also performed for different treatment sites. All maximum passing rates for different categories are 100%. For the brain/skull base, CNS and spine categories, the mean passing rates are 99.1% - 99.3% for FDC and 99.3% - 99.5% for TPS. The TPS is about 0.1% - 0.3% higher than FDC in mean value. The median values are all 100% except CNS for FDC which is 99.9%. The minimum values are within 79.1% - 90.4% for TPS and 80.5% - 88.4% for FDC. The TPS is about - 0.6% - 2.9% higher than FDC in the minimum value.

For the H&N and Lung & Other thoracic treatment sites, the mean passing rates are 98.7% and 98.8% for FDC and TPS is 0.2% lower and 0.1% higher than FDC for the two sites. The median values are 99.6%-100% for both algorithms, and TPS is 0.4% and 0.1% higher than FDC for the two sites respectively. The minimum values are 41.5%, 63.6% for TPS and 59.1%, 62.2% for FDC. TPS is 17.6% lower and 0.6% higher than FDC in the minimum values for the two sites.

For breast cases, the mean passing rate is 95.8% for FDC and 0.8% higher than that of TPS. The minimum passing rate is 57.0% which is 1.5% higher than that of TPS. The median value is 99.0% for TPS and 1.2% higher than that of FDC.

For the GI, pelvic and prostate treatment sites, the mean gamma-index passing rates are 99.5%, 98.9% and 97.4% for TPS and 0.8%, 0.5% and 0.9% higher than that of FDC. The median values are all 100% for TPS and 0.3%, 0.3% and 0.5% higher than that of FDC. The minimum values are 86.7%, 62.5% and 39.4% for TPS and 8.8% higher, 0.7% lower and 2.4% lower than that of FDC for the three sites.

For the “Other” category, both algorithms have the same mean passing rate (97.7%). The TPS has a median value of 99.9%, 0.3% higher than that FDC. FDC has aminimum value of 75.8%, 7.8% higher than that of TPS.

The TPS is slightly better than FDC for Brain/skull base, CNS and spine from the mean/median values (0.1% - 0.3% higher in passing rate). The TPS is better than FDC for GI, pelvic and prostate from the mean values (0.5% - 0.9% higher in passing rate) and the median values (0.3% - 0.5% higher in passing rate). For H&N and breast, the FDC is better from the mean (0.2% and 0.8% higher in passing rate) and minimum values (17.6% and 1.5% higher in passing rate). In the category “Other”, skin, skull, thigh and gluteus patients are included. The TPS and FDC have the same mean passing rate and FDC is better than TPS from the minimum value (7.8% higher in passing rate). In general, the performance of FDC is better for shallower tumours.

Overall, FDC performs marginally below TPS in gamma-index passing rates across most treatment sites; however, its accuracy remains clinically comparable to the TPS-based calculation currently used in practice.

## 4. Discussion

### 4.1 Beam modelling: from single to double Gaussian

The transition from single to double Gaussian spatial and angular distributions in the phase space parameterisation is the central methodological advance of this work, and it is well-motivated by the broader literature. It is well-known that a single-Gaussian beam model cannot accurately predict the low-dose halo made up of particles travelling at large angles to the beam direction, originating from nuclear interactions, inhomogeneous scattering in the nozzle, or large-angle Rutherford scattering. Despite the low magnitude of the halo dose, the large widths of individual halo contributions mean that they may overlap to produce a noticeable impact on the overall dose distribution. The inadequacy of single-Gaussian models for scanning proton beams has been documented since at least Pedroni et al. (2005)[21] and was formally characterised for clinical PBS systems by Zhu et al. (2013)[22], who demonstrated that a double Gaussian fluence model was significantly more accurate than the single Gaussian alternative, though some residual deficiencies in modelling the low-dose envelope remained. Li et al. (2012)[16] showed that low-dose halos in single-spot profiles in a medium could be adequately modelled with the addition of a modified Cauchy–Lorentz component to a double-Gaussian function, and that field-size effects were accurately reproduced at all depths and energies. That the present work achieves 100% gamma-index passing rates for all lateral dose profiles across 20 energies suggests that the double Gaussian parameterisation implemented here is sufficient to capture the halo contribution within the measurement geometry and the 1%/1 mm criterion used, consistent with findings by Kugel et al. (2023), [23]who showed that double Gaussian beam modelling improved agreement between TPS calculations and experimental data in small proton fields, and that MC-based dose engines can benefit from accurate representation of the nozzle spray in the initial beam phase model.

Another distinguishing feature of the present model relative to our previous one for SPHIC is the reorientation of the opening angle distribution center from the beam axis (z-axis) to the line connecting the virtual source to each individual particle's location in the phase space plane. The present approach captures a physically meaningful divergence structure that a purely axial angular distribution would misrepresent, and the 100% lateral profile passing rates confirm that this geometrical correction is effective across the full 70.2–228.7 MeV energy range.

The slightly reduced IDD passing rates at the three lowest energies (97.6% - 98.1% for 70.2 - 89.2 MeV, compared to ≥ 99.0% for higher energies), with deviations concentrated at shallow depth, merit brief comment. The entrance-dose deviation, reaching approximately 4-5% for the lowest three energies, is consistent with the known difficulty of modelling the low-energy halo near the surface. Previous studies[24,23,25] have shown that the low-dose halo contribution becomes increasingly important for small field sizes and low-energy beams, and that inadequate modelling of this component can lead to measurable discrepancies in calculated dose distributions, particularly in the entrance region and for shallow depths.. Since the IDD values reported here represent dose integrated over the full lateral extent of the Giraffe detector (12 cm diameter), the point-dose deviation at the entrance is expected to be considerably smaller than the 4–5% integrated values reported. Nevertheless, this behaviour warrants monitoring during clinical use for shallow, low-energy fields.

*4.2 IMPT plan QA performance in the context of published benchmarks*

The overall mean gamma-index passing rate of 98.5% for FDC and 98.7% for RayStation against 2744 measurements at 3%/3 mm compares favourably with, and in some cases exceeds, those reported for other independent Monte Carlo dose engines validated for clinical proton therapy. Cohilis et al.[26] reported a mean gamma-index passing rate of 99.3% across 62 patient QA planes for an automatically commissioned Monte Carlo engine using 3%/3 mm local criterion, which they noted was well within clinical tolerances. Kaluarachchi et al[27] reported a mean 3%/3 mm passing rate of 99.4% across 160 patient-specific QA measurements for an automated MC framework. Jeon et al. reported [28]absolute gamma-index passing rates of 99.71% at 3%/3 mm for 47 clinical cases using a log-file-based Monte Carlo approach. The present results, while marginally below the highest of these benchmarks, are based on a substantially larger dataset (2744 measurement planes across 269 plans and 155 patients) spanning 10 treatment sites, meaning that more edge cases and challenging anatomical scenarios are captured than in most published comparisons. A broader cohort inherently includes more plans at the tail of the difficulty distribution, which is likely the primary reason for the slightly lower mean passing rate rather than any systematic deficiency in the beam model.

A key methodological strength of the present study is its paired design: FDC and RayStation were applied to the identical set of 269 plans and compared against the same 2744 measurement planes. This means observed differences in gamma-index passing rates between the two engines cannot be attributed to differences in case mix, treatment site distribution, or beam configuration between cohorts - a confound that affects many published comparisons in which independent engines are benchmarked on separate plan sets. The directly paired structure makes even small performance differences between FDC and RayStation more interpretable. In this context, the 0.2% difference in overall mean passing rate (98.5% vs. 98.7%) and the near-identical median values (99.8% vs. 100%) should be understood as reflecting genuine dosimetric similarity between the two engines under matched conditions, rather than as a limitation of FDC. This finding is consistent with the broader literature showing that Monte Carlo-based dose calculation can significantly improve gamma-index passing rates for IMPT patient-specific QA relative to analytical pencil beam algorithms, with MC results showing better or comparable agreement with measurements across most clinical treatment scenarios.

*4.3 Site-specific performance: anatomical and dosimetric explanations*

The site-stratified analysis, made directly interpretable by the paired design, reveals a systematic and clinically meaningful pattern. FDC performs comparably to or better than RayStation for H&N and breast (mean advantage of 0.2% and 0.8% respectively), whereas RayStation holds a modest advantage for GI, pelvic and prostate sites (mean advantage 0.5%–0.9%). Because both algorithms were evaluated on the same plans at the same measurement planes, this gradient cannot reflect differences in cohort composition between the two comparisons — it reflects genuine site-dependent differences in algorithmic performance.

This pattern is consistent with known dosimetric characteristics of the treatment sites involved. For breast and H&N treatments, the shallower tumour location means the beam energy is lower and the relative contribution of the halo to the total dose is greater. As noted above, if the low-energy in-air halo is not fully modelled, this could potentially lead to under-dosing of shallow treatment and the present paired data suggest that FDC's double Gaussian spatial parameterisation captures this contribution more faithfully than RayStation under these conditions.

Conversely, for deep-seated tumours, and prostate in particular, both algorithms encounter longer, more complex beam paths. Prostate cases show the widest spread in passing rates of any site (SD ~7.8–8.4% for both engines), and the lowest minimum values observed in the entire dataset (39.4% for RayStation, 41.8% for FDC). The paired design is especially informative here: the fact that both engines produce their worst results on the same plans confirms that these outliers are driven by plan-specific characteristics - challenging beam geometries - rather than by a systematic weakness of either algorithm. Retrospective review of the specific plans associated with these low outliers would be informative; distinguishing between algorithmic edge cases and delivery-related discrepancies would further clarify the source of these failures and guide future clinical protocol decisions.

The low minimum passing rates observed across several other sites –41.5% for H&N with RayStation and 59.1% for H&N with FDC — also merit attention. The paired structure of the data means these represent the same individual plans evaluated by both engines. The considerably lower RayStation minimum for H&N (17.6% below the corresponding FDC value) is the most pronounced site-specific difference in the dataset, and suggests that for at least some complex H&N plans, the TPS encounters dosimetric scenarios that FDC handles more robustly. Identifying the beam and anatomical characteristics of these specific plans would be a productive avenue for future work.

*4.4 The role of FDC as an independent dose engine*

The clinical and research value of an independent, in-house Monte Carlo engine extends well beyond the QA verification role demonstrated here. Proton centres often develop their own in-house Monte Carlo or analytical second-check systems, given the lack of independent commercial solutions and the absence, until recently, of consensus guidelines for independent dose calculation in proton therapy. The paired validation performed in this study – where FDC and the clinical TPS are simultaneously benchmarked against the same measurement data – provides the strongest possible basis for deploying FDC as a clinical second check tool, since discrepancies between the two engines on future plans can be directly interpreted as genuine dosimetric divergence rather than cohort-level noise. The FDC's track-repeating algorithm, GPU acceleration, and flexibility for implementing non-standard dose quantities – LET, variable RBE models, proton arc optimisation – position it as a research platform that complements rather than replaces the clinical TPS, and the 269 – plan paired dataset validated here lays a rigorous foundation for that expanded use at the enlarged MD Anderson proton centre.

The phase space parameterisation approach used here differs from the source model approach taken by most commercial Monte Carlo implementations. The explicit phase space approach offers greater transparency and transferability between machines but requires careful interpolation for energies not directly commissioned. The linear interpolation of phase space parameters and particle numbers per MU across the 20 measured energies is pragmatic; the accuracy of this interpolation could be further quantified by comparing simulated and measured commissioning data at several interpolated energies, and this would strengthen confidence in the model for clinical energies not directly included in the commissioning set.

*4.5 Limitations*

Several limitations should be acknowledged. First, all measurement comparisons were performed in a homogeneous water-equivalent phantom geometry using the IBA MatriXX PT planar dosimeter array; the accuracy of FDC in heterogeneous patient CT geometry has not been independently validated in this study, although prior FDC publications have addressed this for related systems. Second, the gamma-index analysis at 3%/3 mm, the minimum criterion for IMPT QA per AAPM TG-185 recommendations, is a relatively lenient criterion; reporting of 2%/2 mm results would allow more sensitive discrimination between the two algorithms and enable comparison with a growing body of literature reporting at stricter tolerances. Third, although the paired design is a methodological strength, no formal paired statistical test – such as a Wilcoxon signed-rank test applied to per-measurement passing rates – was performed to determine whether the observed mean differences between FDC and RayStation reach statistical significance. Given the large sample size (2744 measurements), even the small differences observed overall (0.2%) and at the site level (-0.8% – 0.9%) may be statistically significant, and reporting these p-values would meaningfully strengthen the site-specific conclusions. Such analysis is straightforward given the availability of matched per-plane passing rates for both engines and is recommended as a priority revision.

## 5. Conclusion

In this paper, we updated our beam modelling for proton scanning beam and applied it to the University of Texas MD Anderson Cancer Center Proton Therapy Center 2. The updates included using double gaussian function for spatial and opening angle distribution instead of single gaussian function. And also, the center of the angular distribution shifts from along z-axis to particle's original direction. These updates produced the low dose halo over an extended distance from the central axis and signficantly improved the agreement between calculated lateral profiles and measurement.

The number of particles per MU was determined by scaling the calculated absolute point dose to match measurements. Using the generated phase space and MU files, dose calculations were performed for a large cohort of IMPT QA plans and compared with those from the clinically deployed TPS. Gamma-index analysis was used to quantitatively evaluate agreement between calculations and measurements, and the consistently high passing rates confirm the dosimetric accuracy of the commissioned FDC.

## 5. Acknowledgments

Author Qianxia Wang got some help from ChatGPT for abstract simplification due to the 300 word limit for paper submission. Author Pablo Yepes got some help from Claude for the discussion part.

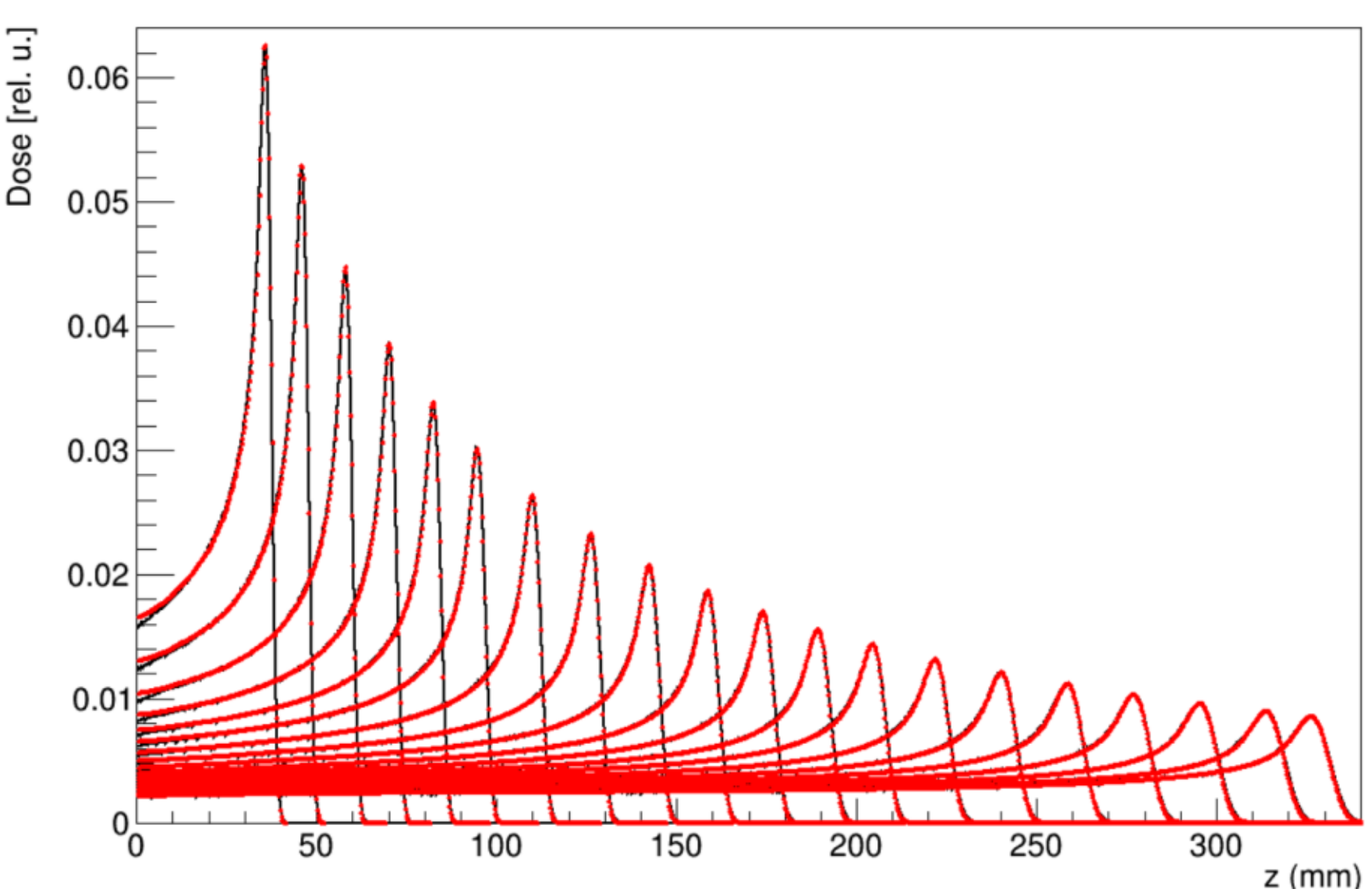

Figure 1. From left to right are intergral depth dose (IDD) for 70.2, 79.2, 89.2, 98.4, 107.0, 115.1, 124.7, 134.3, 143.5, 152.4, 160.3, 168.0, 175.5, 183.7, 192.1, 200.3, 208.3, 216.1, 223.7 and 228.7 MeV/A, respectively. Black solid lines are for FDC simulations and red dot-lines are for measurements.

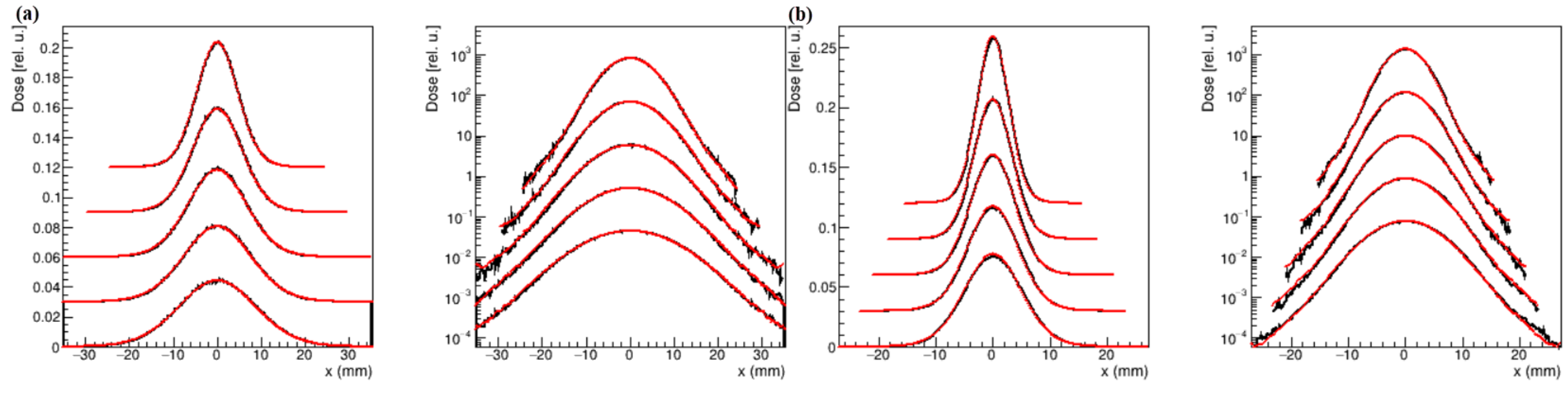

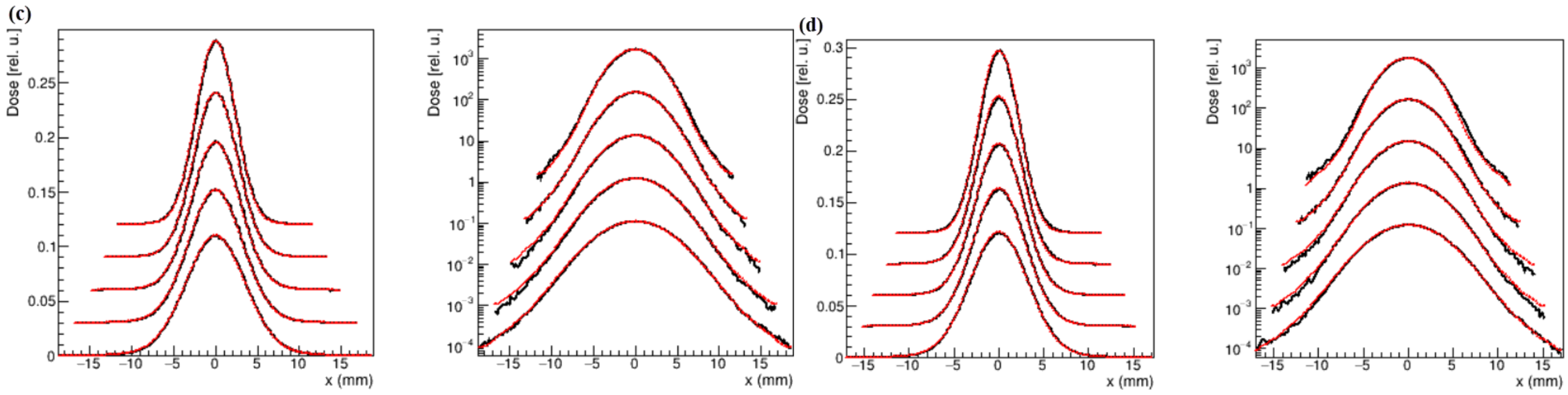

Figure 2. Lateral dose profiles along x axis at 5 depths for four selected energies, 70.2 MeV, 124.7 MeV, 192.1 MeV and 228.7 MeV. The 5 depths are 200 mm, 100 mm, 0 mm, -100 mm and -200 mm from bottom to top with 0.03 up-shift for linear scale (left column) and 10 up-shift for log scales (right column).

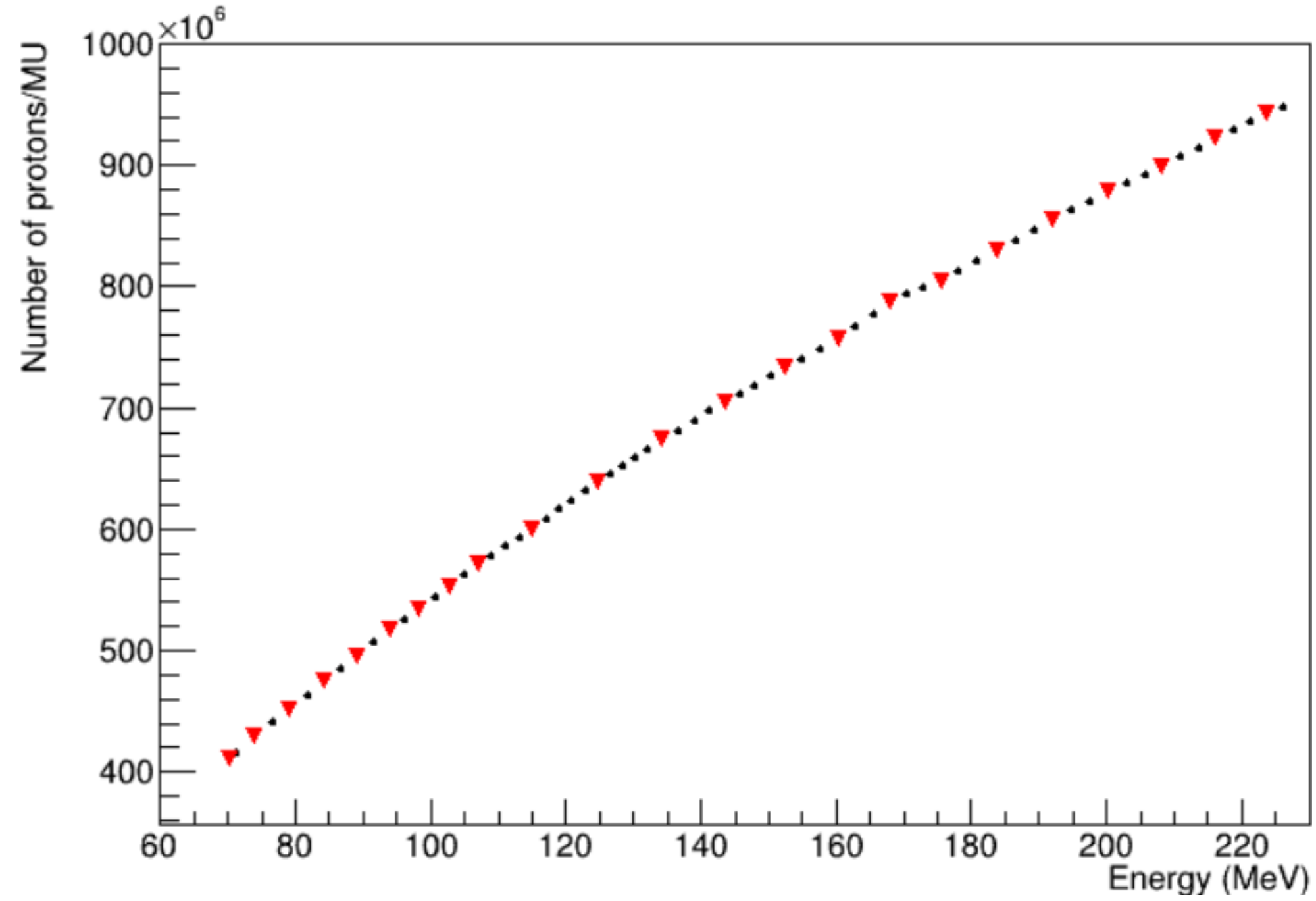

Figure 3. Numbers of protons per MU for different energies. The red triangles are values derived from measurement for the 24 energies: 70.2, 73.9, 79.2, 84.3, 89.2, 93.9, 98.4, 102.7, 107.0, 115.1, 124.7, 134.3, 143.5, 152.4, 160.3, 168.0, 175.5, 183.7, 192.1, 200.3, 208.3, 216.1, 223.7 and 228.7 MeV/A. The black spots are interpolation values for the rest of the available energies.

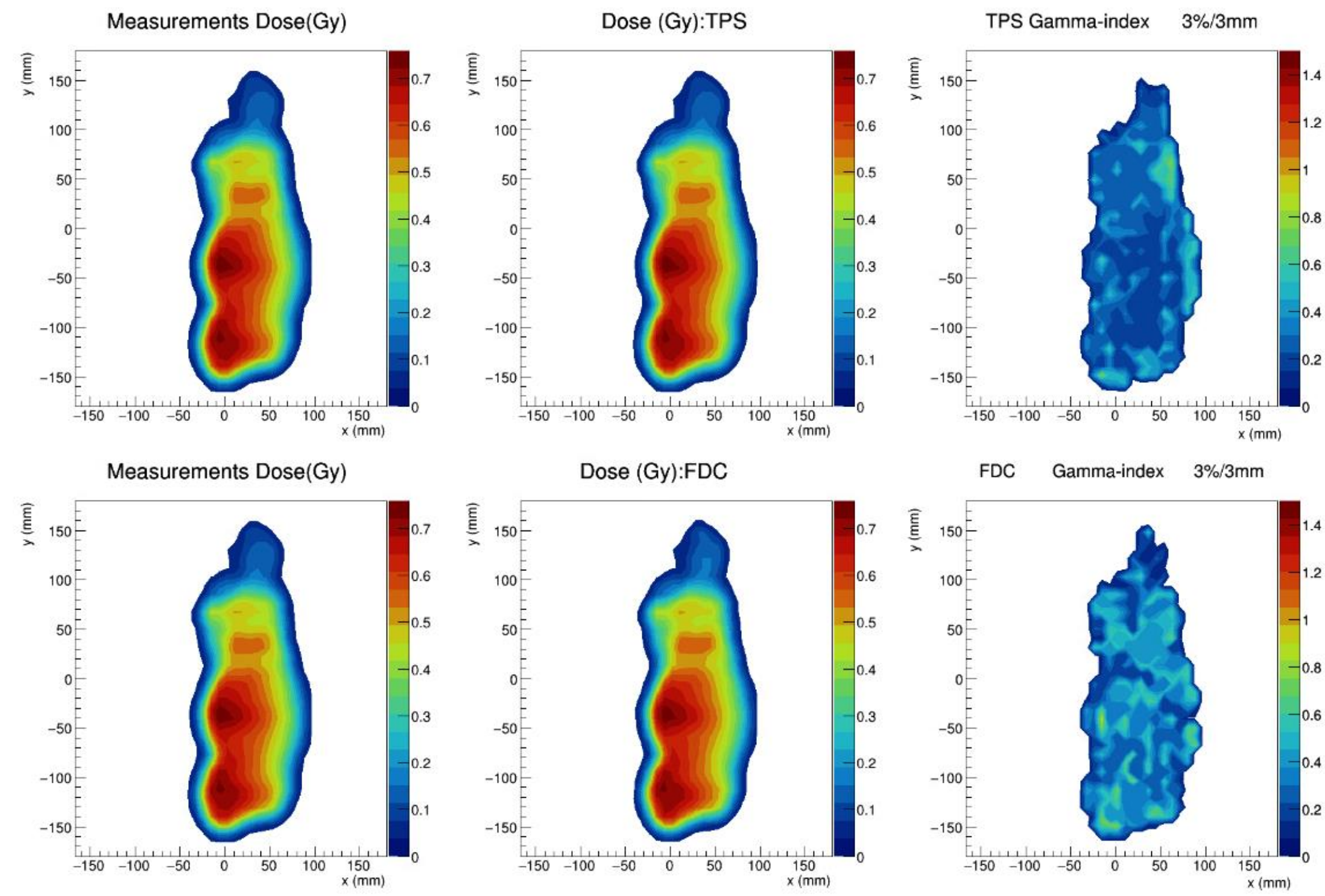

Figure 4 An example of 2D dose measurement (first column), caluclations (second column) and gamma-index distribution.

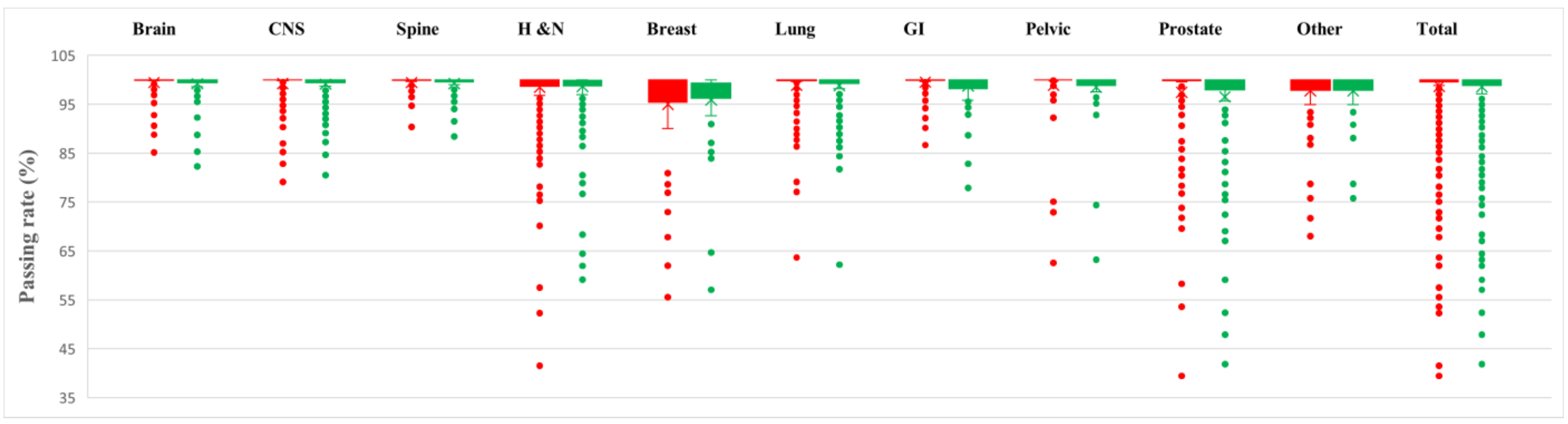

Figure 5. Whisker and box plots for passing rates for different categories and the total patients (red and green are for TPS and FDC).

| Energy (MeV) | IDD (%) | Lateral Dose Profile at 5 depths (%) |
|---|---|---|
| **70.2** | 98.1 | 100 |
| **79.2** | 98.0 | 100 |
| **89.2** | 97.6 | 100 |
| **98.4** | 99.3 | 100 |
| **107.0** | 99.4 | 100 |
| **115.1** | 99.0 | 100 |
| **124.7** | 99.6 | 100 |
| **134.3** | 99.8 | 100 |
| **143.5** | 100 | 100 |
| **152.4** | 100 | 100 |

| **160.3** | 99.7 | 100 |
|---|---|---|
| **168.0** | 99.6 | 100 |
| **175.5** | 100 | 100 |
| **183.7** | 100 | 100 |
| **192.1** | 100 | 100 |
| **200.3** | 100 | 100 |
| **208.3** | 100 | 100 |
| **216.1** | 100 | 100 |
| **223.7** | 100 | 100 |
| **228.7** | 100 | 100 |

Table 1. Gamma-index passing rates for IDD in water and laterlal profiles at different depths (-200 mm, -100 mm, 0 mm, 100 mm and 200 mm) when comparing FDC calculations with measuements with the criterion of 1%/1 mm.

| Site | Engine | # of Patients | # of plans | # of beams | # of measurements | Mean (%) | Median (%) | Min (%) | Max (%) | Δ |
|---|---|---|---|---|---|---|---|---|---|---|
| **Brain & Skull base** | TPS | 16 | 22 | 56 | 270 | 99.5 | 100 | 85.2 | 100 | 1.9 |
| | FDC | | | | | 99.2 | 100 | 82.3 | 100 | 2.1 |
| **CNS** | TPS | 13 | 53 | 105 | 399 | 99.3 | 100 | 79.1 | 100 | 2.5 |
| | FDC | | | | | 99.1 | 99.9 | 80.5 | 100 | 2.3 |
| **Spine** | TPS | 6 | 13 | 32 | 128 | 99.4 | 100 | 90.4 | 100 | 1.6 |
| | FDC | | | | | 99.3 | 100 | 88.4 | 100 | 1.9 |
| **Head & Neck** | TPS | 48 | 70 | 217 | 965 | 98.5 | 100 | 41.5 | 100 | 4.1 |
| | FDC | | | | | 98.7 | 99.6 | 59.1 | 100 | 3.2 |
| **Breast** | TPS | 4 | 8 | 14 | 64 | 95.0 | 99.0 | 55.5 | 100 | 9.4 |
| | FDC | | | | | 95.8 | 97.8 | 57.0 | 100 | 7.3 |
| **Lung & other Thoracic** | TPS | 22 | 29 | 78 | 335 | 98.9 | 100 | 63.6 | 100 | 3.5 |
| | FDC | | | | | 98.8 | 99.9 | 62.2 | 100 | 3.3 |
| **GI** | TPS | 16 | 20 | 43 | 185 | 99.5 | 100 | 86.7 | 100 | 1.6 |
| | FDC | | | | | 98.7 | 99.7 | 77.9 | 100 | 2.6 |
| **Pelvic** | TPS | 6 | 10 | 22 | 104 | 98.9 | 100 | 62.5 | 100 | 5.1 |
| | FDC | | | | | 98.4 | 99.7 | 63.2 | 100 | 5.0 |
| **Prostate** | TPS | 20 | 39 | 55 | 232 | 97.4 | 100 | 39.4 | 100 | 7.8 |
| | FDC | | | | | 96.5 | 99.5 | 41.8 | 100 | 8.4 |
| **Other** | TPS | 4 | 5 | 13 | 62 | 97.7 | 99.9 | 68.0 | 100 | 5.7 |
| | FDC | | | | | 97.7 | 99.6 | 75.8 | 100 | 4.6 |
| **Total** | TPS | 155 | 269 | 635 | 2744 | 98.7 | 100 | 39.4 | 100 | 4.4 |
| | FDC | | | | | 98.5 | 99.8 | 41.8 | 100 | 4.0 |

Table 2 The number of patients, plans, beams and mesurements and gamma-index passing rates analysis for different categories.